\documentclass{article}
\usepackage[dvips]{graphicx}
\topmargin=-1cm\oddsidemargin=0truecm\evensidemargin=0truecm\textheight=24.0cm\textwidth=16.5cm\usepackage{color}
\def\vol#1#2#3{{\bf {#1}} ({#2}) {#3}}\def\PL{Phys.~Lett. }\def\PR{Phys.~Rev. }\def\PRP{Phys.~Rep. }\def\PRL{Phys.~Rev.~Lett. }\def\PTP{Prog.~Theor.~Phys. }\def\IJMP{Int.~J.~Mod.~Phys. }\def\JP{J.~Phys. }

\def\no{\nonumber}

\def\2tvec#1#2{
\left(
\begin{array}{c}
#1  \\
#2  \\   
\end{array}
\right)}

\def\mat2#1#2#3#4{
\left(
\begin{array}{cc}
#1 & #2 \\
#3 & #4 \\
\end{array}
\right)
}

\def\Mat3#1#2#3#4#5#6#7#8#9{
\left(
\begin{array}{ccc}
#1 & #2 & #3 \\
#4 & #5 & #6 \\
#7 & #8 & #9 \\
\end{array}
\right)
}

\def\3tvec#1#2#3{
\left(
\begin{array}{c}
#1  \\
#2  \\   
#3  \\
\end{array}
\right)}

\def\4tvec#1#2#3#4{
\left(
\begin{array}{c}
#1  \\
#2  \\   
#3  \\
#4  \\
\end{array}
\right)}

\def\L{\left}
\def\R{\right}

\def\hbar{\hspace{1mm}\bar{}\hspace{-1mm}h}

\def\eqn#1{
\begin{eqnarray}
#1
\end{eqnarray}
}

\begin{document}

\begin{titlepage}
\begin{flushright}
KIAS-P11062
\end{flushright}
\begin{center}

\vspace{1cm}
{\Large\bf Proton Stability
 in $S_4\times Z_2$ Flavor Symmetric Extra U(1) Model}\\
 \vspace{0.5cm}
 Yasuhiro Daikoku$^{a}$\footnote{yasu\_daikoku@yahoo.co.jp} and
Hiroshi Okada$^{b}$\footnote{hokada@kias.re.kr}
\vspace{5mm}

 {\it 
$^a${ Institute for Theoretical Physics, Kanazawa University, Kanazawa 920-1192, Japan} \\ 
 \vspace{1mm} 
 $^b${School of Physics, KIAS, Seoul 130-722, Korea}\\ \vspace{1mm}
}
  
  \vspace{8mm}

\begin{abstract}
We consider proton stability based on 
$E_6$ inspired extra U(1) model with $S_4\times Z_2$ flavor symmetry.
In this model, a long life time of proton is realized by the flavor symmetry in several ways.
We classify the suppression mechanisms of proton-decay and 
explain how the flavor symmetry works. There is an interesting solution, such as,
in a special direction of vacuum expectation value (VEV),
baryon number violating interactions are canceled.
In the case that the suppression of proton decay is realized by the appropriate size of VEV,
the allowed region of VEV exists when the exotic quark mass is $O(\mbox{TeV})$.
From the constraint for the life time of exotic quark, the right handed neutrino mass should be
in narrow range around $10^{12}\mbox{GeV}$.
\end{abstract}

\end{center}
\end{titlepage}
\setcounter{footnote}{0}

\newpage

\section{Introduction}

Supersymmetry is an elegant solution of hierarchy problem of Standard Model (SM) \cite{SUSY}
and gives a new view point of generation structure of leptons and quarks. 
As a simple supersymmetric extension of SM suffers  from non-conservation of baryon number,
we must introduce R-parity symmetry in order to avoid too rapid proton decay.
This means we can not construct any consistent superpotential based on only SM gauge symmetry.
Even if we introduce R-parity, the superpotential of minimal supersymmetric standard model (MSSM)
is not perfect one. The superpotential of MSSM suffers from $\mu$-problem such as
we must tune the scale of $\mu$-parameter to be $O(\mbox{TeV})$, which is much smaller than Planck scale.
Therefore the R-parity symmetry should be replaced by other symmetry.
The information what symmetry should be introduced may be extracted from the structure of
Yukawa interactions because these interactions are derived from superpotential.

The appropriate start point is given by introducing an additional $U(1)$ gauge symmetry
to forbid $\mu$-term \cite{extra-u1}. 
In this frame work, several new superfields such  as singlet $S$, exotic quarks $G,G^c$,
must be introduced to cancel gauge anomaly.
Then the baryon number violating interactions in superpotential are replaced by  single
exotic quark interactions, which make it easy to suppress proton decay by the new symmetry.

Considering the Yukawa interactions, we can guess about which symmetry we should introduce.
Strangely,  the mixing angle $\theta_{23}$ of Maki-Nakagawa-Sakata (MNS) matrix is almost maximal.
Many authors discussed non-Abelian discrete flavor symmetries to understand the structure of the MNS matrix.
The flavor symmetry may be a good candidate for replacing R-parity symmetry.
At previous work,
we explained $S_4\times Z_2$ flavor symmetry not only realizes maximal mixing angle $\theta_{23}$
but also suppresses proton decay based on
$SU(3)_c\times SU(2)_W\times U(1)_Y\times U(1)_X\times U(1)_Z$ gauge symmetry \cite{s4u1}.
Therefore there is a deep connection between the proton stability and the existence of generation
in supersymmetric model.

In this paper we give more detailed  estimation of proton life time and
classify the mechanisms to suppress proton decay.
At first we give definition of our model in section 2, and give classification of
suppression mechanisms of proton decay in section 3.
In section 4, we check that our superfield assignment realizes 
neutrino mass square differences and MNS matrix including new experimental value of $\theta_{13}$.
We explain the origin of $U(1)_Z$ breaking scale in section 5.
Finally we give conclusion of our analysis in section 6.

\section{$S_4\times Z_2$ flavor symmetric extra U(1) model}

At first we explain the basic structure of our model.
We extend the gauge symmetry to $SU(3)_c\times SU(2)_W\times U(1)_Y\times U(1)_X\times U(1)_Z$
which is the maximal subgroup of $E_6$.
In order to cancel gauge anomaly, we must add new superfields, such as
SM singlet $S$, exotic quark $G,G^c$ (hereafter we call them g-quark) and right handed neutrino (RHN) $N^c$.
We can embed these superfields with MSSM superfields $Q,U^c,D^c,L,E^c,H^U,H^D$
into ${\bf 27}$ of $E_6$ \cite{e6}.
As the singet $S$ develops VEV and breaks $U(1)_X$ gauge symmetry, $O$(TeV) scale $\mu$-term
is induced naturally.
In order to break $U(1)_Z$ and generate a large Majorana mass of RHN,
we add SM singlet $\Phi,\Phi^c$.
The gauge representations of superfields are given in Table 1 \cite{s4u1}.
\begin{table}[htbp]
\begin{center}
\begin{tabular}{|c|c|c|c|c|c|c|c|c|c|c|c||c|c|}
\hline
         &$Q$ &$U^c$    &$E^c$&$D^c$    &$L$ &$N^c$&$H^D$&$G^c$    &$H^U$&$G$ &$S$ &$\Phi$&$\Phi^c$\\ \hline
$SU(3)_c$&$3$ &$3^*$    &$1$  &$3^*$    &$1$ &$1$  &$1$  &$3^*$    &$1$  &$3$ &$1$ &$1$   &$1$     \\ \hline
$SU(2)_W$&$2$ &$1$      &$1$  &$1$      &$2$ &$1$  &$2$  &$1$      &$2$  &$1$ &$1$ &$1$   &$1$     \\ \hline
$y=6Y$   &$1$ &$-4$     &$6$  &$2$      &$-3$&$0$  &$-3$ &$2$      &$3$  &$-2$&$0$ &$0$   &$0$     \\ \hline
$x$      &$1$ &$1$      &$1$  &$2$      &$2$ &$0$  &$-3$ &$-3$     &$-2$ &$-2$&$5$ &$0$   &$0$     \\ \hline
$z$      &$-1$&$-1$     &$-1$ &$2$      &$2$ &$-4$ &$-1$ &$-1$     &$2$  &$2$ &$-1$&$8$   &$-8$    \\ \hline
$R$      &$-$ &$-$      &$-$  &$-$      &$-$ &$-$  &$+$  &$+$      &$+$  &$+$ &$+$ &$+$   &$+$     \\ \hline
\end{tabular}
\end{center}
\caption{$G_2$ assignment of superfields.
Where $x$, $y$ and $z$ are charges of $U(1)_X$, $U(1)_Y$ and $U(1)_Z$, respectively. $Y$ is hypercharge.
After the gauge symmetry breaking of three $U(1)$s, R-parity symmetry
$R=\exp[i\pi(3x-8y+15z)/20]$ is unbroken.
}
\end{table}

Under the gauge symmetry given in Table 1, the renormalizable superpotential is given by
\eqn{
W&=& Y^UH^UQU^c +Y^DQD^cH^D +Y^EH^DLE^c +Y^NH^ULN^c +Y^M\Phi N^cN^c+\lambda SH^UH^D+kSGG^c \no \\
&+&M\Phi \Phi^c +Y^{QQ}GQQ +Y^{UD}G^cU^cD^c +Y^{UE}GE^cU^c +Y^{LQ}G^cLQ +Y^{ND}GN^cD^c.
}
In this superpotential, unwanted terms are included in the second line.
The first term of the second line is the mass term of singlets $\Phi, \Phi^c$ which prevent singlets
from developing VEVs. The other five terms of the second line are single g-quark interactions, which
break baryon and lepton number and induce rapid proton decay.
In the first line, we must take care of the flavor changing neutral currents (FCNCs)
induced by extra Higgs bosons \cite{e6-FCNC}.
Therefore the superpotential Eq.(1) is not consistent at the present stage.

In order to stabilize proton, we introduce $S_4\times Z_2$ flavor symmetry.
If we assign $G,G^c$ to $S_4$ triplet and quarks and leptons to doublet or singlet,
the single g-quark interaction is forbidden. 
However, as the g-quark must never be stable from phenomenological reason, 
we assign $\Phi^c$ to $S_4$ triplet to break the flavor symmetry slightly.
In order to realize the maximal mixing angle of $\theta_{23}$ in the MNS matrix and
suppress the Higgs-mediated FCNCs, we assign the superfields 
in our model as given in Table 2 \cite{s4pamela}.

In the non-renormalizable part of superpotential, 
the single g-quark interactions which contribute to the g-quark decay are given as follows 
\eqn{
W_B&=&\frac{y^{QQ}}{M^2_P}\Phi\Phi^cQQG
+\frac{y^{UD}}{M^2_P}\Phi\Phi^cG^cU^cD^c
+\frac{y^{EU}}{M^2_P}\Phi\Phi^cGE^cU^c
+\frac{y^{QL}}{M^2_P}\Phi\Phi^cG^cLQ .
}
The detail of $W_B$ depends on the $Z_2$ charge assignment of $p_q$ and $p_g$.

\begin{table}[htbp]
\begin{center}
\begin{tabular}{|c|c|c|c|c|c|c|c|c|c|}
\hline
        &$Q_1$      &$Q_2$      &$Q_3$       &$U^c_1$    &$U^c_2$    &$U^c_3$     &$D^c_1$    
&$D^c_2$    &$D^c_3$ \\
      \hline
$S_4$   &${\bf 1}$  &${\bf 1}$  & ${\bf 1}$  &${\bf 1}$  &${\bf 1}$  &${\bf 1}$   &${\bf 1}$
&${\bf 1}$  &${\bf 1}$\\
      \hline
$Z_2$   &$p_q$      &$p_q$      &$p_q$       &$p_q$      &$p_q$      &$p_q$       &$p_q$      
&$p_q$      &$p_q$ \\
      \hline
      \hline
        &$E^c_1$  &$E^c_2$  &$E^c_3$   &$L_i$    &$L_3$    &$N^c_i$   &$N^c_3$  &$H^D_i$  &$H^D_3$  \\
      \hline
$S_4$   &${\bf 1}$&${\bf 1}$&${\bf 1'}$&${\bf 2}$&${\bf 1}$&${\bf 2}$ &${\bf 1}$&${\bf 2}$&${\bf 1}$ \\
      \hline      
$Z_2$   &$+$      &$-$      &$+$       &$-$      &$-$      &$+$       &$-$      &$-$      &$+$ \\
      \hline
      \hline   
        &$H^U_i$  &$H^U_3$  &$S_i$     &$S_3$    &$G_a$      &$G^c_a$   &$\Phi_i$ &$\Phi_3$ &$\Phi^c_a$\\
      \hline
$S_4$   &${\bf 2}$&${\bf 1}$&${\bf 2}$ &${\bf 1}$&${\bf 3}$  &${\bf 3}$ &${\bf 2}$&${\bf 1}$&${\bf 3}$\\
      \hline
$Z_2$   &$-$      &$+$      &$-$       &$+$      &$p_g$      &$p_g$     &$+$      &$+$      &$+$ \\
      \hline
\end{tabular}
\end{center}
\caption{$S_4\times Z_2$ assignment of superfields
(Where the index $i$ of the $S_4$ doublets runs $i=1,2$,
and the index $a$ of the $S_4$ triplets runs $a=1,2,3$. The charges of quarks and g-quarks; $p_q$ and
$p_g$, take $\pm$.)}
\end{table}

\section{Classification of the suppression mechanism of proton decay}

Depending on the $p_q$ and $p_g$, the flavor symmetry works in different ways
to suppress proton decay. 
In this section we classify the suppression mechanism and estimate the
allowed parameter range.\\
{\bf (1)Leptoquark solution}

If we assign $(p_q,p_g)=(+,-)$, it results
$y^{QQ}=y^{UD}=0$ in Eq.(2).
In this case we can assign lepton number $(L)$ and baryon number $(B)$ of $G$ to $(L,B)=(+1,1/3)$ and
those of $G^c$ to $(L,B)=(-1,-1/3)$, then the baryon number is conserved.
Therefore proton becomes stable  and VEVs of $\Phi,\Phi^c$ are not bounded from above. 
In this solution, $G,G^c$ are well known as leptoquarks.
Note that the VEVs of $\Phi,\Phi^c$ are bounded from bellow, because
the life time of g-quark must be shorter than 0.1 sec, otherwise the success of BBN is spoiled
\cite{g-lifetime}. 

If we assign $(p_q,p_g)=(-,-)$, then it results $y^{QQ}=y^{UD}=y^{QL}=0$ and
g-quark becomes leptoquark which couples only to right handed charged leptons $E^c_1,E^c_3$.
Therefore the decay mode of our leptoquark depends on the flavor charge assignment.\\
{\bf (2)Cancellation solution}

If we assign $(p_q,p_g)=(+,+)$, it results $y^{QL}=0$. In this case, as we cannot
define the lepton number and baryon number of g-quark, these numbers are not conserved.
Here we investigate proton decay interactions which are induced by scalar g-quarks exchange.
In the present flavor assignment, the superpotential which contributes to the proton decay is given by
\eqn{
W_B&=&\frac{y_a}{M^2_P}U^c_aE^c_3
[\sqrt{3}(G_2\Phi^c_2-G_3\Phi^c_3)\Phi_2-(G_2\Phi^c_2+G_3\Phi^c_3-2G_1\Phi^c_1)\Phi_1] \no \\
&+&\frac{y_{ab}}{M^2_P}Q_aQ_b\Phi_3(\Phi^c_1G_1+\Phi^c_2G_2+\Phi^c_3G_3)
+\frac{y'_{ab}}{M^2_P}U^c_aD^c_b\Phi_3(\Phi^c_1G^c_1+\Phi^c_2G^c_2+\Phi^c_3G^c_3) \no \\
&+&\frac{z_{ab}}{M^2_P}Q_aQ_b
[\sqrt{3}(G_2\Phi^c_2-G_3\Phi^c_3)\Phi_1+(G_2\Phi^c_2+G_3\Phi^c_3-2G_1\Phi^c_1)\Phi_2] \no \\
&+&\frac{z'_{ab}}{M^2_P}U^c_aD^c_b
[\sqrt{3}(G^c_2\Phi^c_2-G^c_3\Phi^c_3)\Phi_1+(G^c_2\Phi^c_2+G^c_3\Phi^c_3-2G^c_1\Phi^c_1)\Phi_2].
}
Note  that the contribution from $E^c_1$ is omitted  because it is shown to be $\tau$ lepton
in section 4 and does not  contribute to proton decay.
 Integrating out scalar g-quarks, we get the effective four-Fermi interactions
as follows
\eqn{
{\cal L}&=&\frac{C_{GG}}{M^4_PM^2_G}\sum_{abc}\lambda_{abc}\mu^c u^c_a\bar{q}_b\bar{q}_c
+\frac{C_{GG^c}}{M^4_PM^2_G}\sum_{abc}\lambda'_{abc}\mu^c u^c_au^c_bd^c_c, \no \\
\lambda_{abc}
&=&y_ay_{bc}\L[\sqrt{3}\L<\Phi_3\R>\L<\Phi_2\R>\L(\L<\Phi^c_2\R>^2-\L<\Phi^c_3\R>^2\R)
+\L<\Phi_3\R>\L<\Phi_1\R>\L(2\L<\Phi^c_1\R>^2-\L<\Phi^c_2\R>^2-\L<\Phi^c_3\R>^2\R)\R] \no \\
&+&y_az_{bc}\L[2\L<\Phi_1\R>\L<\Phi_2\R>\L(\L<\Phi^c_2\R>^2+\L<\Phi^c_3\R>^2-2\L<\Phi^c_1\R>^2\R)
-\sqrt{3}\L(\L<\Phi_1\R>^2-\L<\Phi_2\R>^2\R)\L(\L<\Phi^c_2\R>^2-\L<\Phi^c_3\R>^2\R)\R], \no \\
\lambda'_{abc}
&=&y_ay'_{bc}\L[\sqrt{3}\L<\Phi_3\R>\L<\Phi_2\R>\L(\L<\Phi^c_2\R>^2-\L<\Phi^c_3\R>^2\R)
+\L<\Phi_3\R>\L<\Phi_1\R>\L(2\L<\Phi^c_1\R>^2-\L<\Phi^c_2\R>^2-\L<\Phi^c_3\R>^2\R)\R] \no \\
&+&y_az'_{bc}\L[2\L<\Phi_1\R>\L<\Phi_2\R>\L(\L<\Phi^c_2\R>^2+\L<\Phi^c_3\R>^2-2\L<\Phi^c_1\R>^2\R)
-\sqrt{3}\L(\L<\Phi_1\R>^2-\L<\Phi_2\R>^2\R)\L(\L<\Phi^c_2\R>^2-\L<\Phi^c_3\R>^2\R)\R],
}
where $\mu^c=e^c_3$ and mean scalar g-quark mass $M_G$ and dimensionless coefficients 
$C_{GG}, C_{GG^c}$ are defined in appendix.
Note that the masses of scalar g-quarks are degenerated due to the $S_4$ symmetry.
Interestingly, it results $\lambda_{abc}=\lambda'_{abc}=0$ in the special VEV direction such as
\eqn{
\L<\Phi^c_1\R>=\L<\Phi^c_2\R>=\L<\Phi^c_3\R> .
}
In this case proton decay is forbidden. This means the contributions from three 
scalar g-quarks are canceled. \\
{\bf (3)Suppression solution}

In the case that there is no remarkable cancellation, the size of $\L<\Phi\R>$ must 
be in appropriate region where the constraints for proton and g-quark life time 
are satisfied at the same time \cite{king}.
In order to suppress proton decay, the VEV of $\Phi$ must not be too large.
At first we estimate the upper bound of the VEV.
As there are many unknown parameters in $W_B$ and the VEV direction of $\Phi, \Phi^c$ is also unknown,
we make several assumption for simplicity.
At first, we assume there is no mixing between scalar g-quarks $G$ and $G^c$ and put 
$C_{GG}=1,C_{GG^c}=0$. Next, we change the assignment 
of $Q,U^c,D^c$  and $G,G^c$ to $(S_4,Z_2)=({\bf 1'}, +)$ and $(S_4,Z_2)=({\bf 3},+)$ 
respectively in Table 2 and replace the superpotial in Eq.(3) by
\eqn{
W_B&=&\frac{y_a}{M^2_P}U^c_aE^c_3
[\sqrt{3}(G_2\Phi^c_2-G_3\Phi^c_3)\Phi_1+(G_2\Phi^c_2+G_3\Phi^c_3-2G_1\Phi^c_1)\Phi_2] \no \\
&+&\frac{z_a}{M^2_P}U^c_aE^c_3\Phi_3(G_1\Phi^c_1+G_2\Phi^c_2+G_3\Phi^c_3) \no \\
&+&\frac{y_{ab}}{M^2_P}Q_aQ_b\Phi_3(G_1\Phi^c_1+G_2\Phi^c_2+G_3\Phi^c_3) \no \\
&+&\frac{z_{ab}}{M^2_P}Q_aQ_b
[\sqrt{3}(G_2\Phi^c_2-G_3\Phi^c_3)\Phi_1+(G_2\Phi^c_2+G_3\Phi^c_3-2G_1\Phi^c_1)\Phi_2],
}
where we assume $y_a=z_{ab}=0$ and the contribution from $y^{UD}$ is omitted. 
Finally we tune the VEV direction as follows
\eqn{
\L<\Phi^c_1\R>=\L<\Phi^c_2\R>=\L<\Phi^c_3\R>
=\L<\Phi_1\R>=\L<\Phi_2\R>=\L<\Phi_3\R>=\frac{V}{\sqrt{3}}.
}
Including the renormalization factor $A_{RF}$, the effective four-Fermi interactions at 1GeV is
given by \cite{p-RGE}
\eqn{
{\cal L}&=&\sum_{abc}\frac{A_{RF}z_ay_{bc}V^4}{3M^4_PM^2_G}\mu^cu^c_a\bar{q}_b\bar{q}_c, \\
A_{RF}&=&(A^y_{RF})_S(A^z_{RF})_S(A_{RF})_L ,
}
where we estimate the short distance part of $A_{RF}$ by the 1-loop renormalization group
equations as follows
\eqn{
(4\pi)\frac{d\ln z_a}{d\ln\mu}&=&-\frac{16}{3}\alpha_s ,\\
(4\pi)\frac{d\ln y_{ab}}{d\ln\mu}&=&-\frac{24}{3}\alpha_s.
}
Here only QCD correction is accounted. This approximation is not bad
because the beta function of the coupling constant of strong interaction $g_s$ vanishes at 1-loop level
in our model, which makes the contribution of $\alpha_s$ dominant in  the RGEs of $z_a$ and $y_{ab}$.
Solving Eq.(10) and Eq.(11), if we put $\alpha_s(M_Z)=0.118$, we get
\eqn{
(A^z_{RF})_S&=&\L(\frac{M_P}{M_Z}\R)^{4\alpha_s/3\pi}
=\L(\frac{2.43\times 10^{18}}{91}\R)^{0.05008}=6.647, \\
(A^y_{RF})_S&=&\L(\frac{M_P}{M_Z}\R)^{2\alpha_s/\pi}
=\L(\frac{2.43\times 10^{18}}{91}\R)^{0.07512}=17.139.
}
The long distance part is given by \cite{p-decay}
\eqn{
(A_{RF})_L=\L(\frac{\alpha_s(1\mbox{GeV})}{\alpha_s(m_b)}\R)^{6/25}
\L(\frac{\alpha_s(m_b)}{\alpha_s(M_Z)}\R)^{6/23}=1.4,
}
from which we get
\eqn{
A_{RF}=159.5.
}
For the case that the final state includes $\mu^+$,
the strongest experimental bound for the partial decay width of proton is given by 
$p\to\pi^0+\mu^+$.
For simplicity, we assume $z_1=z_2=z_3=1$ for the mass eigenstates $u^c_a$ and
tune $y_{bc}$ to normalize four-Fermi interaction as follows
\eqn{
{\cal L}_{eff}=\L(\frac{V}{M_P}\R)^4\frac{A_{RF}}{M^2_G}\bar{u}\bar{d}u^c\mu^c .
}
From this Lagrangian, the proton decay width is given by \cite{PDform}
\eqn{
\Gamma(p\to \pi^0+\mu^+)&=&\frac{m_p}{64\pi f^2_\pi}
\L[\L(\frac{V}{M_P}\R)^4\frac{A_{RF}}{M^2_G}\R]^2(1+F+D)^2
\L(1-\frac{m^2_{\pi^0}}{m^2_p}\R)^2\alpha^2_p ,  
}
where $F$ and $D$ are chiral Lagrangian parameters,
$\alpha_p$ is hadronic matrix element, $f_\pi$ is pion decay constant and
$m_{\pi^0}$ and $m_p$ are masses of pion and proton.
If we put
\eqn{
F=0.47,\quad D=0.80,\quad \alpha_p=-0.012\ \mbox{GeV}^3,\quad f_\pi=130\ \mbox{MeV}, \quad
m_{\pi^0}=135\ \mbox{MeV},\quad m_p=940\ \mbox{MeV} \cite{chiral},
}
then we get
\eqn{
\Gamma(p\to \pi^0+\mu^+)&=&(5.01 \times 10^{-12}GeV)
\L[\L(\frac{V}{M_P}\R)^4\frac{(1000\ \mbox{GeV})^2}{M^2_G}\R]^2 .
}
From the experimental bound $\tau(p\to \pi^0+\mu^+)>473\times 10^{30}[\mbox{years}]$
\cite{PDG}, the upper bound for VEV is estimated as follows
\eqn{
\L[\L(\frac{V}{M_P}\R)^4\L(\frac{1000\ \mbox{GeV}}{M_G}\R)^2\R]^2<8.78 \times 10^{-54}.
}
Next, we estimate the life time of g-quark under the assumption that g-quark 
is lighter than scalar g-quark.
For simplicity, we assume the g-quark can decay only into smuon but not into stau or squarks.
With this assumption, the g-quarks decay through the following interaction
\eqn{
{\cal L}=\frac{(A^z_{RF})_SV^2}{3M^2_P}(u^c+c^c+t^c)
\tilde{\mu}^c(g_1+g_2+g_3) ,
}
from which one can see that g-quarks have the same life time.
Requiring the life time of $g_a$ is shorter than 0.1 sec as follows
\eqn{
\Gamma(g_a)=3\L(\frac{(A^z_{RF})_SV^2}{3M^2_P}\R)^2\frac{M_g}{16\pi}
>\frac{1}{0.1\ \mbox{sec}} , 
}
we get
\eqn{
\frac{M_g}{1000 \mbox{GeV}}\L(\frac{V}{M_P}\R)^4>2.25 \times 10^{-26},
}
where $M_g$ is g-quark mass. Hereafter we assume the approximation $M_g=M_G$ is held for simplicity.
From Eq.(20) and Eq.(23), the allowed region for $V$ is given by (see Fig.1)
\eqn{
2.25\times 10^{-26}\L(\frac{1000 \mbox{GeV}}{M_G}\R)<\L(\frac{V}{M_P}\R)^4
<2.96\times 10^{-27}\L(\frac{M_G}{1000\ \mbox{GeV}}\R)^2 .
}
This inequality holds when the mass bound, 
\eqn{
M_G> 1.96\ \mbox{TeV},
}
is satisfied. For example, if we put $M_G=10$ TeV, allowed region for $V$ is given by
\eqn{
0.53<\frac{V}{10^{12}\ \mbox{GeV}}<1.79 .
}
Note that the factor of this constraint should not be taken seriously,
because there is large model dependence.
\begin{figure}[ht]
\unitlength=1mm \hspace{3cm}
\begin{picture}(70,70)
\includegraphics[height=6cm,width=10cm]{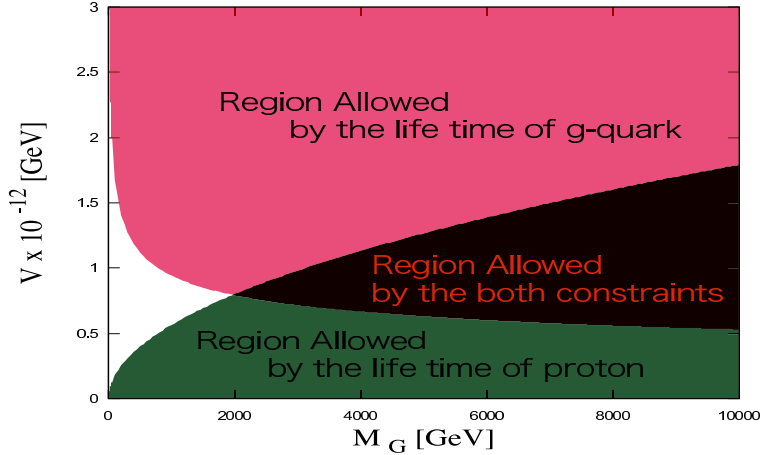}
\end{picture}
\caption{ $M_G$ versus $V$: The pink region comes from the constraint of the life time of the g-quark, which should be less than 0.1 sec.  The green region comes from the constraint of the proton stability. The black region is allowed by the both constraints. The heavier of $M_G$, the wider the allowed region is.}
\label{proton}
\end{figure}

\section{The Maki-Nakagawa-Sakata matrix}

In this section we confirm the assignment of Table 2
realizes the neutrino masses and MNS matrix.
In the superpotential
\eqn{
W_L&=&Y^N_2\L[H^U_1(L_1N^c_2+L_2N^c_1)+H^U_2(L_1N^c_1-L_2N^c_2)\R] \no \\
&+&Y^N_3H^U_3L_3N^c_3+Y^N_4L_3(H^U_1N^c_1+H^U_2N^c_2) \no \\
&+&Y^E_1E^c_1(H^D_1L_1+H^D_2L_2)+Y^E_2E^c_2H^D_3L_3+Y^E_3E^c_3(H^D_1L_2-H^D_2L_1) \no \\
&+&\frac12 Y^M_1\Phi_3(N^c_1N^c_1+N^c_2N^c_2)+ \frac12 Y^M_3\Phi_3 N^c_3N^c_3 \no \\
&+&\frac12Y^M_2[2\Phi_1N^c_1N^c_2+\Phi_2(N^c_1N^c_1-N^c_2N^c_2)],
}
we define the VEVs of scalar fields as follows
\eqn{
&&\L<H^U_1\R>=\L<H^U_2\R>=\frac{1}{\sqrt{2}}v_u,\quad 
\L<H^U_3\R>=v'_u ,\quad 
\L<H^D_1\R>=\L<H^D_2\R>=\frac{1}{\sqrt{2}}v_d,\quad 
\L<H^D_3\R>=v'_d ,\no \\
&&\L<\Phi_1\R>=|a|V_0c_N,\quad \L<\Phi_2\R>=|a|V_0s_N,\quad
\L<\Phi_3\R>=V_0=\frac{V}{\sqrt{1+|a|^2}},
}
and define the mass parameters as follows \cite{kubo}
\eqn{
\begin{tabular}{llll}
$M_1=Y^M_1V_0$,        & $M_3=Y^M_3V_0$,          & $M_2=Y^M_2|a|V_0$     &   \\
$m^\nu_2=Y^N_2v_u$,  & $m^\nu_3=|Y^N_3|v'_u$, & $m^\nu_4=Y^N_4v_u$, &   \\
$m^l_1=Y^E_1v_d$,    & $m^l_2=Y^E_2v'_d$,     & $m^l_3=Y^E_3v_d$.   &
\end{tabular}
}
Without loss of generality, by the field redefinition, we can define
$Y^E_{1,2,3},Y^M_{1,3},Y^N_{2,4}$ are real and non-negative and
$Y^M_2,Y^N_3$ are complex. For simplicity, we put
\eqn{
Y^M_1=Y^M_3=1,\quad Y^M_2=e^{i\psi},
}
and
\eqn{
a=e^{i\psi}|a|,\quad M_1=M_3=V_0,\quad M_2=aV_0.
}
With these parameters, the mass matrices are given by
\eqn{
\begin{tabular}{ll}
$M_l=\frac{1}{\sqrt{2}}\Mat3{m^l_1}{0}{-m^l_3}{m^l_1}{0}{m^l_3}{0}{\sqrt{2}m^l_2}{0}$, &
$M_D=\frac{1}{\sqrt{2}}\Mat3{m^\nu_2}{m^\nu_2}{0}{m^\nu_2}{-m^\nu_2}{0}
{m^\nu_4}{m^\nu_4}{\sqrt{2}e^{i\delta}m^\nu_3}$,  \\
$M_R=V_0\Mat3{1+as_N}{ac_N}{0}{ac_N}{1-as_N}{0}{0}{0}{1}$. &
\end{tabular}
}
Due to the seesaw mechanism, the neutrino mass matrix is given by
\eqn{
M_\nu&=&M_DM^{-1}_RM^t_D=\frac{1}{1-a^2}
\Mat3{\rho^2_2(1-ac_N)}{-\rho^2_2as_N}{\rho_2\rho_4(1-ac_N)}
{-\rho^2_2as_N}{\rho^2_2(1+ac_N)}{-\rho_2\rho_4as_N}
{\rho_2\rho_4(1-ac_N)}{-\rho_2\rho_4as_N}{\rho^2_4(1-ac_N)+\rho^2_3(1-a^2)}, \no  \\
&&\rho_2=\frac{m^\nu_2}{\sqrt{V_0}},\quad
\rho_4=\frac{m^\nu_4}{\sqrt{V_0}},\quad
\rho_3=\frac{e^{i\delta}m^\nu_3}{\sqrt{V_0}}.
}
The charged lepton mass matrix is diagonalized as follows
\eqn{
V^\dagger_l M^*_l M^t_l V_l&=&diag(m^2_e,m^2_\mu, m^2_\tau)=((m^l_2)^2,(m^l_3)^2,(m^l_1)^2), \\
V_l&=&\frac{1}{\sqrt{2}}\Mat3{0}{-1}{1}
{0}{1}{1}
{-\sqrt{2}}{0}{0}.
}
To realize experimental results \cite{neutrino},
the neutrino mass matrix should be diagonalized as follows
\eqn{
V^t_\nu M_\nu V_\nu&=&diag(m_{\nu_1},m_{\nu_2},m_{\nu_3}) , \no \\
V_{MNS}&=&V^\dagger_lV_\nu
=\Mat3{-c_\nu c_{13}}{-s_\nu c_{13}}{s_{13}}
{\frac{1}{\sqrt{2}}s_\nu+\frac{1}{\sqrt{2}}c_\nu s_{13}}
{-\frac{1}{\sqrt{2}}c_\nu+\frac{1}{\sqrt{2}}s_\nu s_{13}}
{\frac{1}{\sqrt{2}}c_{13}}
{-\frac{1}{\sqrt{2}}s_\nu+\frac{1}{\sqrt{2}}c_\nu s_{13}}
{\frac{1}{\sqrt{2}}c_\nu+\frac{1}{\sqrt{2}}s_\nu s_{13}}
{\frac{1}{\sqrt{2}}c_{13}} , \no \\
&&\sin^2 2\theta_\nu=0.8704 ,\quad
\sin^2 2\theta_{13}=0.11 \no \\
&&m^2_{\nu_2}-m^2_{\nu_1}=7.6\times 10^{-5}\ [\mbox{eV}^2] ,\quad
m^2_{\nu_3}-m^2_{\nu_2}=2.4\times 10^{-3}\ [\mbox{eV}^2].
}
Due to the overabundance of parameters, unfortunately,
it is impossible to fix the parameters by these constraints.
Therefore we assume $a$ and $\rho^2_3$  are real for simplicity, then we get
\eqn{
&&a=1.40321 , \quad s_N=0.110194 ,\no \\
&&\rho^2_2=0.0212781\ [\mbox{eV}] , \quad \rho^2_3=-0.0539146\ [\mbox{eV}] , \quad
\rho^2_4=0.112142\ [\mbox{eV}] , \\
&&m_{\nu_1}=-0.0207554\ [\mbox{eV}] ,\quad m_{\nu_2}= 0.0225119\ [\mbox{eV}] ,\quad
m_{\nu_3}=-0.0539146\ [\mbox{eV}] , \no
}
and
\eqn{
V=\sqrt{1+|a|^2}\frac{(Y^N_4v_u)^2}{\rho^2_4}=(1.54\times 10^{12})\L(Y^N_4\frac{v_u}{10\ \mbox{GeV}}\R)^2\ [\mbox{GeV}] .
}
From the requirement of perturbativity of Yukawa coupling
such as $Y^N_4<1$, $V$ is bounded from above. 
It is difficult for $V$ to be much larger than $O(10^{12}\mbox{GeV})$.
Therefore, there exists upper bound for $V$
even in the case that proton decay is perfectly forbidden.
It must be noted that there is non-trivial
coincidence of the constraint of Eq.(26) with RHN mass scale.

Finally we give a comment about how to realize the VEV directions given in Eq.(28).
The Higgs potential derived from superpotential
\eqn{
W_H&=&\lambda_1S_3(H^U_1H^D_1+H^U_2H^D_2)+\lambda_3S_3H^U_3H^D_3 \no \\
&+&\lambda_4H^U_3(S_1H^D_1+S_2H^D_2)+\lambda_5(S_1H^U_1+S_2H^U_2)H^D_3,
}
has accidental $O(2)$ symmetry. To avoid massless Nambu-Goldstone boson,
we add soft $S_4\times Z_2$ breaking terms in the form of the inner products with $(1,1)$ as follows
\eqn{
{\cal L}\supset m^2_{BU}(H^U_3)^\dagger(H^U_1+H^U_2)
+m^2_{BD}(H^D_3)^\dagger(H^D_1+H^D_2)
+m^2_{BS}(S_3)^\dagger(S_1+S_2)+h.c. ,
}
then the VEV direction $(A_1,A_2)\propto (1,1)$ $(A=S,H^U,H^D)$ becomes
the minimum of potential
and the VEV direction of $H^U_a,H^D_a,S_a$ in Eq.(28) is realized.
The potential of $\Phi_a,\Phi^c_a$ derived from the leading order superpotential
\eqn{
W_\Phi&=&\frac{A_1}{2M_P}\Phi^2_3\L[(\Phi^c_1)^2+(\Phi^c_2)^2+(\Phi^c_3)^2\R] \no \\
&+&\frac{A_2}{2M_P}(\Phi^2_1+\Phi^2_2)\L[(\Phi^c_1)^2+(\Phi^c_2)^2+(\Phi^c_3)^2\R] \no \\
&+&\frac{A_3}{2M_P}\L\{2\sqrt{3}\Phi_1\Phi_2\L[(\Phi^c_2)^2-(\Phi^c_3)^2\R]
+(\Phi^2_1-\Phi^2_2)\L[(\Phi^c_2)^2+(\Phi^c_3)^2-2(\Phi^c_1)^2\R]\R\} \no \\
&+&\frac{A_4}{2M_P}\Phi_3\L\{\sqrt{3}\Phi_1\L[(\Phi^c_2)^2-(\Phi^c_3)^2\R]
+\Phi_2\L[(\Phi^c_2)^2+(\Phi^c_3)^2-2(\Phi^c_1)^2\R]\R\} ,
}
does not have accidental symmetry.
To avoid domain wall problem, we must add soft $S_4$ breaking terms. 
Then the VEV direction of $\Phi,\Phi^c$ is controlled 
by the parameters $A_{1,2,3,4}$ and soft SUSY and flavor breaking parameters.
The mechanism for inducing soft flavor symmetry breaking terms is unknown and
beyond the scope of this paper. We leave it for future work.

\section{The origin of the scale of $V$}

Finally  we explain how the required value for $V$ is realized.
The superpotential Eq.(41) is simplified as follows
\eqn{
W_\Phi=\frac{A}{M_P}(\Phi\Phi^c)^2.
}
The origin of  $\Phi$-potential becomes unstable point due to the negative
soft SUSY breaking squared mass and the potential is lifted by
F-term derived by $W_\Phi$. Minimizing the potential, the VEV of $\Phi$
is estimated as follows
\eqn{
V\sim \L<\Phi\R>=\L(\frac{m_{SUSY}M_P}{A}\R)^\frac12 .
}
For the typical range of $A$ and SUSY breaking scale $m_{SUSY}$ such  as
$0.01<A<1,\ 0.1\ \mbox{TeV}<m_{SUSY}<10\ \mbox{TeV}$. Hence the region of $V$ is as follows
\eqn{
10^{10}\ \mbox{GeV}<V<10^{12}\ \mbox{GeV} .
}
Although the each of region given in Eq.(26) and Eq.(44) is very narrow, remarkably, 
there exists overlap.

\section{Conclusion}

We have considered  the suppression mechanism of proton decay
based on $S_4\times Z_2$ flavor symmetric model.
Under the field assignment that MNS matrix is realized,
we have classified the several suppression mechanisms.
There are two new solutions
other than the well known leptoquark solution.
For the cancellation solution, the four-Fermi interaction
which induces proton decay vanishes in a special VEV direction.
For the suppression solution, the stability of proton is satisfied
for appropriate size of VEV. 
Although the allowed region for the VEV is very narrow,
there is coincidence between the allowed
regions required by the different phenomenological considerations such as
naive potential analysis, RHN mass scale and
the life times of g-quark and proton.

\appendix

\section{Mixing matrix of scalar g-quarks}

Here we define the mixing matrix of scalar g-quarks.
The mass terms of scalar g-quarks are given as follows
\eqn{
-{\cal L} &\supset& m^2_G(|G_1|^2+|G_2|^2+|G_3|^2)+m^2_{G^c}(|G^c_1|^2+|G^c_2|^2+|G^c_3|^2) \no \\
&+&kA_k[S_3(G_1G^c_1+G_2G^c_2+G_3G^c_3)+h.c.] \no \\
&+&\L|k(G_1G^c_1+G_2G^c_2+G_3G^c_3)+\lambda_1(H^U_1H^D_1+H^U_2H^D_2)+\lambda_3H^U_3H^D_3\R|^2\no \\
&+&|k|^2|S_3|^2(|G_1|^2+|G_2|^2+|G_3|^2+|G^c_1|^2+|G^c_2|^2+|G^c_3|^2)+\mbox{D-terms} \no \\
&=&\sum_a(G^*_a,G^c_a)\mat2{M^2_{++}}{M^2_{+-}}{M^2_{+-}}{M^2_{--}}\2tvec{G_a}{(G^c_a)^*},
}
where we assumed $A_k$ is real for simplicity. If
\eqn{
M^2_{+-}=kA_kv'_s+k(\lambda_1v_uv_d+\lambda_3v'_uv'_d)=0,
}
is satisfied, then there is no $G-G^c$ mixing. In the case that $M^2_{+-}\neq 0$,
the mixing matrix of scalar g-quarks is defined as follows
\eqn{
V_G=\mat2{c_G}{-s_G}{s_G}{c_G},\quad
V^T_G\mat2{M^2_{++}}{M^2_{+-}}{M^2_{+-}}{M^2_{--}}V_G=\mbox{diag}(M^2_+,M^2_-),\quad
\2tvec{G_a}{(G^c_a)^*}=V_G\2tvec{G_{+,a}}{G_{-,a}}.
}
From this definition, the propagators of scalar g-quarks are given by
\eqn{
\L<G_a,G^*_b\R>&=&\delta_{ab}\L(\frac{c^2_G}{M^2_+}+\frac{s^2_G}{M^2_-}\R)
=\delta_{ab}\frac{C_{GG}}{M^2_G},\quad C_{GG}=c^2_G\frac{M^2_G}{M^2_+}+s^2_G\frac{M^2_G}{M^2_-}, \no \\
\L<G_a,G^c_b\R>&=&\delta_{ab}c_Gs_G\L(\frac{1}{M^2_+}-\frac{1}{M^2_-}\R)
=\delta_{ab}\frac{C_{GG^c}}{M^2_G},\quad C_{GG^c}=c_Gs_G\L(\frac{M^2_G}{M^2_+}-\frac{M^2_G}{M^2_-}\R),
} 
where $M_G=\sqrt{M_+M_-}$ is mean scalar g-quark mass.



\begin{thebibliography}{99}
\bibitem{SUSY}
H.~P.~Nilles, \PRP\vol{110}{1984}{1}.

\bibitem{extra-u1}D.~Suematsu and Y.~Yamagishi, \IJMP\vol{A10}{1995}{4521}.

\bibitem{s4u1}Y.~Daikoku and H.~Okada,
 \PR\vol{D82}{2010}{033007}[arXiv:0910.3370[hep-ph]].

\bibitem{e6}F.~Zwirner, \IJMP\vol{A3}{1988}{49},
J.~L.~Hewett and T.~G.~Rizzo, \PRP\vol{183}{1989}{193}.

\bibitem{e6-FCNC}B.~A.~Campbell, J.~Ellis, K.~Enqvist, M.~K.~Gaillard and D.~V.~Nanopoulos,
\IJMP\vol{A2}{1987}{831};
  Y.~Daikoku and H.~Okada,
  [arXiv:1008.0914 [hep-ph]].



\bibitem{s4pamela}Y.~Daikoku, H.~Okada and T.~Toma,
\PTP {\bf126} (2011) 855-883 [arXiv:1106.4717 hep-ph]].

\bibitem{g-lifetime}M.~Kawasaki, K.~Kohri and T.~Moroi, \PR\vol{D71}{2005}{083502}
[astro-ph/0408426].


\bibitem{king}R.~Howl and S.~F.~King, JHEP\vol{0805}{2008}{008}[arXiv:0802.1909[hep-ph]].

\bibitem{p-RGE}J.~Hisano,
[hep-ph/0004266].

\bibitem{p-decay}P.~Nath, and P.~F.~Perez,
\PRP\vol{441}{2007}{191} [hep-ph/0601023].

\bibitem{PDform} T.~Goto and T.~Nihei, 
\PR\vol{D59}{1999}{115009}[hep-ph/9808255].

\bibitem{chiral}Y.~Aoki, C.~Dawson, J.~Noaki, and A.~Soni,
\PR\vol{D75}{2007}{014507}[hep-lat/0607002].

\bibitem{PDG}Particle Data Group, \JP\vol{G37}{2010}{075021}
and 2011 partial update for the 2012 edition .


\bibitem{kubo}J.Kubo,
\PL\vol{B578}{2004}{156}.

\bibitem{neutrino}T2K Collaboration: K.~Abe et.al, \PRL\vol{107}{2011}{041801}[arXiv:11062822[hep-ex]].
\end{thebibliography}
\end{document}